\documentclass[twocolumn]{aastex631}
\usepackage{graphicx}   
\usepackage{lipsum}

\shorttitle{The CAMELS project: public data release}
\shortauthors{The CAMELS team}
\graphicspath{{./}{figures/}}

\begin{document}

\title{The CAMELS project: public data release}

\correspondingauthor{The CAMELS team}
\email{camel.simulations@gmail.com}

\author[0000-0002-4816-0455]{Francisco Villaescusa-Navarro}
\affiliation{Center for Computational Astrophysics, Flatiron Institute, 162 5th Avenue, New York, NY, 10010, USA}
\affiliation{Department of Astrophysical Sciences, Princeton University, Peyton Hall, Princeton NJ 08544, USA}

\author{Shy Genel}
\affiliation{Center for Computational Astrophysics, Flatiron Institute, 162 5th Avenue, New York, NY, 10010, USA}
\affiliation{Columbia Astrophysics Laboratory, Columbia University, New York, NY, 10027, USA}

\author[0000-0001-5769-4945]{Daniel Angl\'es-Alc\'azar}
\affiliation{Department of Physics, University of Connecticut, 196 Auditorium Road, Storrs, CT, 06269, USA}
\affiliation{Center for Computational Astrophysics, Flatiron Institute, 162 5th Avenue, New York, NY, 10010, USA}

\author{Lucia A. Perez}
\affiliation{Arizona State University, School of Earth and Space Exploration, 781 Terrace Mall Tempe, AZ 85287, USA}

\author[0000-0002-0936-4279]{Pablo Villanueva-Domingo}
\affiliation{Instituto de F\'isica Corpuscular (IFIC), CSIC-Universitat de Val\`encia, E-46980, Paterna, Spain}

\author{Digvijay Wadekar}
\affiliation{Center for Cosmology and Particle Physics, Department of Physics, New York University, New York, NY 10003, USA}
\affiliation{School of Natural Sciences, Institute for Advanced Study, 1 Einstein Drive, Princeton, NJ 08540, USA}

\author{Helen Shao}
\affiliation{Department of Astrophysical Sciences, Princeton University, Peyton Hall, Princeton NJ 08544, USA}

\author[0000-0001-9243-7434]{Faizan G. Mohammad}
\affiliation{Waterloo Center for Astrophysics, University of Waterloo, Waterloo, ON N2L 3G1, Canada}
\affiliation{Department of Physics and Astronomy, University of Waterloo, Waterloo, ON N2L 3G1, Canada}

\author{Sultan Hassan}
\affiliation{Center for Computational Astrophysics, Flatiron Institute, 162 5th Avenue, New York, NY, 10010, USA}
\affiliation{Department of Physics \& Astronomy, University of the Western Cape, Cape Town 7535, South Africa}

\author[0000-0003-1593-1505]{Emily Moser}
\affiliation{Department of Astronomy, Cornell University, Ithaca, NY 14853, USA}

\author{Erwin T. Lau}
\affiliation{Center for Astrophysics | Harvard \& Smithsonian, 60 Garden St., Cambridge, MA 02138, USA}

\author[0000-0002-1948-3562]{Luis Fernando Machado Poletti Valle}
\affiliation{Institute for Particle Physics and Astrophysics, ETH Z\"urich, Wolfgang-Pauli-Strasse 27, CH-8093 Z\"urich, Switzerland}

\author{Andrina Nicola}
\affiliation{Department of Astrophysical Sciences, Princeton University, Peyton Hall, Princeton NJ 08544, USA}

\author[0000-0003-2911-9163]{Leander Thiele}
\affiliation{Department of Physics, Princeton University, Princeton, NJ 08544, USA}

\author{Yongseok Jo}
\affiliation{Center for Theoretical Physics, Department of Physics and Astronomy, Seoul National University, Seoul 08826, Korea}

\author{Oliver H.\,E. Philcox}
\affiliation{Department of Astrophysical Sciences, Princeton University, Peyton Hall, Princeton NJ 08544, USA}
\affiliation{School of Natural Sciences, Institute for Advanced Study, 1 Einstein Drive, Princeton, NJ 08540, USA}

\author{Benjamin D. Oppenheimer}
\affiliation{CASA, Department of Astrophysical and Planetary Sciences, University of Colorado, 389 UCB, Boulder, CO 80309, USA}
\affiliation{Center for Astrophysics | Harvard \& Smithsonian, 60 Garden St., Cambridge, MA 02138, USA}

\author[0000-0002-1185-4111]{Megan Tillman}
\affiliation{Department of Physics and Astronomy, Rutgers University, 136 Frelinghuysen Road, Piscataway, NJ 08854, USA} 

\author[0000-0003-1197-0902]{ChangHoon Hahn}
\affiliation{Department of Astrophysical Sciences, Princeton University, Peyton Hall, Princeton NJ 08544, USA}

\author[0000-0003-4786-2348]{Neerav Kaushal}
\affiliation{Department of Physics, Michigan Technological University, Houghton, MI 49931, USA}

\author[0000-0002-6146-4437]{Alice Pisani}
\affiliation{Center for Computational Astrophysics, Flatiron Institute, 162 5th Avenue, New York, NY, 10010, USA}
\affiliation{The Cooper Union for the Advancement of Science and Art, 41 Cooper Square, New York, NY 10003, USA}
\affiliation{Department of Astrophysical Sciences, Princeton University, Peyton Hall, Princeton NJ 08544, USA}

\author{Matthew Gebhardt}
\affiliation{Department of Physics, University of Connecticut, 196 Auditorium Road, Storrs, CT, 06269, USA}

\author{Ana Maria Delgado}
\affiliation{Center for Astrophysics | Harvard \& Smithsonian, 60 Garden St., Cambridge, MA 02138, USA}

\author{Joyce Caliendo}
\affiliation{Department of Physics, University of Connecticut, 196 Auditorium Road, Storrs, CT, 06269, USA}
\affiliation{Department of Astronomy, University of Massachussets Amherst, Amherst, MA 01003, USA}

\author{Christina Kreisch}
\affiliation{Department of Astrophysical Sciences, Princeton University, Peyton Hall, Princeton NJ 08544, USA}

\author{Kaze W.K. Wong}
\affiliation{Center for Computational Astrophysics, Flatiron Institute, 162 5th Avenue, New York, NY, 10010, USA}

\author{William R. Coulton}
\affiliation{Center for Computational Astrophysics, Flatiron Institute, 162 5th Avenue, New York, NY, 10010, USA}

\author{Michael Eickenberg}
\affiliation{Center for Computational Mathematics, Flatiron Institute, 162 5th Avenue, New York, NY, 10010, USA}

\author[0000-0002-2539-2472]{Gabriele Parimbelli}
\affiliation{Dipartimento di Matematica e Fisica. Universit\'a Roma Tre, via della Vasca Navale 84, I--00146, Roma, Italy.}
\affiliation{INFN -- National Institute for Nuclear Physics, via della Vasca Navale 84, I--00146 Roma, Italy}
\affiliation{Scuola Internazionale Superiore di Studi Avanzati.  via Bonomea, 265, I--34136, Trieste, Italy}
\affiliation{INAF-OATs, Osservatorio Astronomico di Trieste, Via Tiepolo 11, I--34131 Trieste, Italy.}
\affiliation{IFPU -- Institute for Fundamental Physics of the Universe, Via Beirut 2, I--34151 Trieste, Italy.}

\author{Yueying Ni}
\affiliation{McWilliams Center for Cosmology, Department of Physics, Carnegie Mellon University, Pittsburgh, PA 15213, USA}

\author[0000-0001-8867-5026]{Ulrich P. Steinwandel}
\affiliation{Center for Computational Astrophysics, Flatiron Institute, 162 5th Avenue, New York, NY, 10010, USA}

\author{Valentina La Torre}
\affiliation{Department of Physics and Astronomy, Tufts University, Medford, MA 02155, USA}

\author{Romeel Dave}
\affiliation{Institute for Astronomy, University of Edinburgh, Royal Observatory, Edinburgh EH9 3HJ, UK}
\affiliation{Department of Physics \& Astronomy, University of the Western Cape, Cape Town 7535, South Africa}
\affiliation{South African Astronomical Observatories, Observatory, Cape Town 7925, South Africa}

\author{Nicholas Battaglia}
\affiliation{Department of Astronomy, Cornell University, Ithaca, NY 14853, USA}

\author{Daisuke Nagai}
\affiliation{Department of Physics, Yale University, New Haven, CT 06520, USA}

\author{David N. Spergel}
\affiliation{Center for Computational Astrophysics, Flatiron Institute, 162 5th Avenue, New York, NY, 10010, USA}
\affiliation{Department of Astrophysical Sciences, Princeton University, Peyton Hall, Princeton NJ 08544, USA}

\author{Lars Hernquist}
\affiliation{Center for Astrophysics | Harvard \& Smithsonian, 60 Garden St., Cambridge, MA 02138, USA}

\author{Blakesley Burkhart}
\affiliation{Department of Physics and Astronomy, Rutgers University, 136 Frelinghuysen Road, Piscataway, NJ 08854, USA} 
\affiliation{Center for Computational Astrophysics, Flatiron Institute, 162 5th Avenue, New York, NY, 10010, USA}

\author{Desika Narayanan}
\affiliation{Department of Astronomy, University of Florida, Gainesville, FL, USA}
\affiliation{University of Florida Informatics Institute, 432 Newell Drive, CISE Bldg E251, Gainesville, FL, USA}

\author{Benjamin Wandelt}
\affiliation{Sorbonne Universite, CNRS, UMR 7095, Institut d’Astrophysique de Paris, 98 bis boulevard Arago, 75014 Paris, France}
\affiliation{Center for Computational Astrophysics, Flatiron Institute, 162 5th Avenue, New York, NY, 10010, USA}

\author{Rachel S. Somerville}
\affiliation{Center for Computational Astrophysics, Flatiron Institute, 162 5th Avenue, New York, NY, 10010, USA}

\author{Greg L. Bryan}
\affiliation{Department of Astronomy, Columbia University, 550 W 120th Street, New York, NY 10027, USA}
\affiliation{Center for Computational Astrophysics, Flatiron Institute, 162 5th Avenue, New York, NY, 10010, USA}

\author[0000-0002-2642-5707]{Matteo Viel}
\affiliation{Scuola Internazionale Superiore di Studi Avanzati.  via Bonomea, 265, I--34136, Trieste, Italy}
\affiliation{IFPU -- Institute for Fundamental Physics of the Universe, Via Beirut 2, I--34151 Trieste, Italy.}
\affiliation{INAF-OATs, Osservatorio Astronomico di Trieste, Via Tiepolo 11, I--34131 Trieste, Italy.}
\affiliation{INFN -- National Institute for Nuclear Physics, Via Valerio 2, I-34127 Trieste, Italy}

\author[0000-0002-0701-1410]{Yin Li}
\affiliation{Center for Computational Astrophysics, Flatiron Institute, 162 5th Avenue, New York, NY, 10010, USA}
\affiliation{Center for Computational Mathematics, Flatiron Institute, 162 5th Avenue, New York, NY, 10010, USA}

\author{Vid Irsic}
\affiliation{Kavli Institute for Cosmology, University of Cambridge, Madingley Road, Cambridge CB3 0HA, UK}
\affiliation{Cavendish Laboratory, University of Cambridge, 19 J. J. Thomson Ave., Cambridge CB3 0HE, UK}

\author{Katarina Kraljic}
\affiliation{Aix Marseille Univ, CNRS, CNES, LAM, Marseille, France}

\author{Mark Vogelsberger}
\affiliation{Kavli Institute for Astrophysics and Space Research, Department of Physics, MIT, Cambridge, MA 02139, USA}

\begin{abstract}

The Cosmology and Astrophysics with MachinE Learning Simulations (CAMELS) project was developed to combine cosmology with astrophysics through thousands of cosmological hydrodynamic simulations and machine learning. 
CAMELS contains 4,233 cosmological simulations, 2,049 N-body and 2,184 state-of-the-art hydrodynamic simulations that sample a vast volume in parameter space. In this paper we present the CAMELS public data release, describing the characteristics of the CAMELS simulations and a variety of data products generated from them, including halo, subhalo, galaxy, and void catalogues, power spectra, bispectra, Lyman-$\alpha$ spectra, probability distribution functions, halo radial profiles, and X-rays photon lists. We also release over one thousand catalogues that contain billions of galaxies from CAMELS-SAM: a large collection of N-body simulations that have been combined with the Santa Cruz Semi-Analytic Model. We release all the data, comprising more than 350 terabytes and containing 143,922 snapshots, millions of halos, galaxies and summary statistics. We provide further technical details on how to access, download, read, and process the data at \url{https://camels.readthedocs.io}. 

\end{abstract}


\section{Introduction} 
\label{sec:intro}

Recent advances in deep learning are triggering a revolution across fields, and cosmology and astrophysics are not left behind. Applications include parameter inference \citep{Paco_2021a, Paco_2021b, Siamak_16, Cole_2021, Michelle_2019, Peek_2019, Mangena_2020, Hassan_2020}, superresolution \citep{Doogesh_2020, Yin_2020a, Yueying_2021}, generation of mock data \citep{Hassan_2021, DW_2021, Zamudio_2019}, painting hydrodynamic properties on N-body simulations \citep{Zhang_2019, Jacky_2019, Renan_2020, Noah_2020, Jay_2019, Leander_2020, Jo_2019, Horowitz_2021, Harrington_2021,Bernardini2022,Moews_2021}, improving the halo-galaxy connection \citep{Moster_2021, Jay_2020, Xu_2021, Lovell_2021, Del21}, removing/cleaning astrophysical effects \citep{Lucas_2020,  Pablo_2020, Liu_2021}, emulating non-linear evolution and speeding up numerical simulations \citep{He_19, NECOLA, Doogesh_2019, 2021arXiv211205681D}, learning functions to interpolate among simulation properties \citep{Giusarma_19, Chen_2020}, estimating masses of dark matter halos \citep{2019ApJ...881...74M, 2019MNRAS.490.2367C, 2020arXiv201110577L, GNN_CAMELS, GNN_MW_M31} and galaxy clusters \citep{Ntampaka:2018rjt, Ho:2019zap, Ramanah:2020ift, Ramanah:2020ylz,2020MNRAS.499.3445Y,2020ApJ...900..110G,deAndres:2021tjl}, finding universal relations in subhalo properties \citep{Shao_2021}, generating realistic galaxy images \citep{2019MNRAS.485.3203F}, model selection and classification~\citep[e.g.][]{Hassan_2019}, and improving SED fitting techniques \citep{lovell_2019,gilda_2020}, among many others (see \citealt{george_stein_2020_4024768}\footnote{\url{https://github.com/georgestein/ml-in-cosmology}} for a comprehensive compilation). At its core, many of these results are based on using neural networks to approximate complex functions that may live in a high dimensional space. These techniques have the potential to revolutionize the way we do cosmology and astrophysics. 

From the cosmological side we have now a well established and accepted model: the $\Lambda$ cold dark matter ($\Lambda$CDM) model. This model not only describes the laws and constituents of our Universe, but it is also capable of explaining a large variety of cosmological observables, from the temperature anisotropies of the cosmic microwave background to the spatial distribution of galaxies at low redshift. The model has free parameters characterizing fundamental properties of the Universe such as its geometry, composition, the properties of dark energy, the sum of the neutrino masses, etc. One of the most important tasks in cosmology is to constrain the values of the these parameters with the highest degree of accuracy. In that way, we may be able to provide answers to fundamental questions such as: ``What is the nature of dark energy?'' and ``What are the masses of the neutrinos?''

Many studies have shown that there is a wealth of cosmological information located on mildly to highly non-linear scales that need summary statistics other than the power spectrum to be retrieved \citep{Quijote, Samushia_2021, Gualdi_2021, Kuruvilla_2021, Bayer_2021,  Banerjee_2019, Changhoon_2019, Uhlemann_2020, Friedrich_2020, Massara_2020, Dai_2020, Allys_2020, Banerjee_2020, Banerjee_2021, Gualdi_2020, Gualdi_2021, Giri_2020, Bella_2020, Changhoon_2020, Valgiannis_2021, Bayer_2021, Kuruvilla_2021b, Naidoo_2021, Porth_2021, Hortua_2021}. Extracting the maximum amount of information from these scales presents two main challenges. First, the optimal summary statistics that fully characterizes non-Gaussian density fields is currently unknown. Second, these scales are expected to be affected by astrophysical effects, such as feedback from supernovae and active galactic nuclei (AGN), in a poorly understood way \citep[e.g.][]{SomervilleDave2015, Naab_2017}. Due to this uncertainty, cosmological analysis are typically carried out avoiding scales that are affected by astrophysical processes. 

On the other hand, the cosmological dependence on astrophysical processes such as the formation and evolution of galaxies is typically neglected. Thus, while intrinsically linked, cosmology and galaxy formation tend to progress in parallel with limited interactions. Building bridges between cosmology and galaxy formation will thus benefit the development of both branches and contribute to an unified understanding.

Unfortunately, the interplay of cosmology and astrophysics takes places on many different scales, including non-linear ones. This implies that cosmological hydrodynamic simulations are among the best tools to model and study the interactions between cosmology and astrophysics. However, given the uncertainties in both cosmology and galaxy formation models, it would be desirable to run simulations for different values of the cosmological parameters and also for different astrophysical models. Finally, if the number of simulations is large enough, one can make use of machine learning techniques to extract the maximum amount of information from the simulations while at the same time being able to develop high-dimensional interpolators to explore the parameter space without having to run additional simulations.

The Cosmology and Astrophysics with MachinE Learning Simulations (CAMELS) project \citep{CAMELS} was conceived to combine cosmology and astrophysics through numerical simulations and machine learning. At its core, CAMELS consists of a set of 4,233 cosmological simulations that have different values of the cosmological parameters and different astrophysical models. All these virtual universes can be used as a large dataset to train machine learning algorithms.

The CAMELS project was first introduced and described in detail in \citet{CAMELS}. The theoretical justification behind some of its main features (e.g. the use of a latin-hypercube covering a big volume in parameter space) was presented in \citet{Villaescusa-Navarro_2020c}. Since then, a number of different works have made use of the CAMELS simulations to carry out a large and diverse variety of tasks:
\begin{enumerate}
\item In \cite{Shao_2021} CAMELS was used to identify a universal relation between subhalo properties using neural networks and symbolic regression. 
\item In \cite{Faizan_2021} CAMELS was used to train convolutional neural networks to inpaint masked regions of highly non-linear 2D maps from different physical fields. 
\item In \cite{Paco_2021a} CAMELS was used to show that neural networks can extract cosmological information and marginalize over baryonic effects at the field level using multiple fields simultaneously.
\item In \cite{Paco_2021b} CAMELS was used to show that neural networks can place robust, percent level, constraints on $\Omega_{\rm m}$ and $\sigma_8$ from 2D maps containing the total matter mass of hydrodynamic simulations.
\item In \cite{Paco_2021c} the CAMELS Multified Dataset, a collection of hundreds of thousands of 2D maps and 3D grids for 13 different fields was presented and publicly released.
\item In \cite{Hassan_2021} CAMELS was used to train a generative model that can produce diverse neural hydrogen maps by end of reionization (z$\sim$6) as a function of cosmology.
\item In \cite{GNN_CAMELS}, a model based on Graph Neural Networks (GNNs) was trained on the data from the CAMELS simulations to predict the total mass of a dark matter halo given its galactic properties while accounting for astrophysical uncertainties.
\item In \cite{GNN_MW_M31} the GNN models proposed in \cite{GNN_CAMELS} and trained on CAMELS data were used to obtain the first constrain on the mass of the Milky Way and Andromeda using artificial intelligence.
\item In \cite{Nicola_2022} CAMELS was used to investigate the potential of auto- and cross-power spectra of the baryon distribution to robustly constrain cosmology and baryonic feedback.
\item In \cite{Wad21} CAMELS was used to reduce the scatter in the Sunyaev-Zeldovich (SZ) flux-mass relation, $Y$-$M$, to provide more accurate estimates of cluster masses.
\item In \cite{WadThi22} CAMELS was used to study deviations from self-similarity in the $Y$-$M$ relation due to baryonic feedback processes, and to find an alternative relation which is more robust.
\item In \cite{Thiele_2022} CAMELS was used to demonstrate the strong constraints that next-generation measurements of the $y$-distortions could provide on feedback models.
\item In \citet{Moser22} CAMELS was used to compute thermal and kinetic SZ profiles. A Fisher analysis was performed to forecast the constraining power of observed SZ profiles on the astrophysical models varied in the simulations.
\item In \cite{Paco_2022a} CAMELS was used to investigate whether the value of the cosmological parameters can be constrained using properties of a single galaxy.
\item In \citet{Jo22} CAMELS has been exploited to infer the full posterior on the combinations of cosmological and astrophysical parameters that reproduce observations such as cosmic star formation history and stellar mass functions using simulation-based inference. 
\item \citet{Lucia_2022} created CAMELS-SAM, a third larger `hump' of CAMELS by combining N-body simulations with the Santa Cruz semi-analytic model of galaxy formation. CAMELS-SAM contains billions of galaxies and represents a perfect tool to investigate and quantify the amount of cosmological information that can be extracted with galaxy redshift surveys.
\end{enumerate}
In this paper we describe the characteristics of the CAMELS simulations together with a variety of data products obtained from them, and we publicly release all available data. This paper is accompanied by the online documentation hosted at \url{https://camels.readthedocs.io}, containing further technical details on how to access, read, and manipulate CAMELS data. We believe that the CAMELS data will trigger new developments and findings in the fields of cosmology and galaxy formation.

This paper is organized as follows. In Sec.~\ref{sec:simulations} we briefly describe the simulations of the CAMELS project and their scientific goals. The specifications of the data release are outlined in detail in Sec.~\ref{sec:data}. In Sec.~\ref{sec:access} we describe how to access and download the data together with the overall data organization. We conclude in Sec.~\ref{sec:summary}.

\section{Simulations} 
\label{sec:simulations}

\subsection{Overview}

CAMELS consists of a set of 4,233 cosmological simulations: 2,049 N-body and 2,184 hydrodynamic. All simulations follow the evolution of $256^3$ dark matter particles and $256^3$ fluid elements (only the hydrodynamic simulations) from $z=127$ down to $z=0$ in a periodic box of $(25~h^{-1}{\rm Mpc})^3$ volume. The initial conditions were generated at $z=127$ using second order perturbation theory (2LPT)\footnote{\url{https://cosmo.nyu.edu/roman/2LPT/}}. The linear power spectra were computed using \textsc{CAMB} \citep{CAMB}. The mass resolution is approximately $1.27\times10^7~h^{-1}M_\odot$ per baryonic resolution element and the gravitational softening length is approximately 2~${\rm kpc}$. For each simulation we have saved 34 snapshots, from $z=6$ to $z=0$. All simulations share the value of these cosmological parameters: $\Omega_{\rm b}=0.049$, $h=0.6711$, $n_s=0.9624$, $\sum m_\nu=0.0$ eV, $w=-1$. However, the value of $\Omega_{\rm m}$ and $\sigma_8$ varies from simulation to simulation. 
\begin{figure*}
\begin{center}
\includegraphics[width=0.99\linewidth]{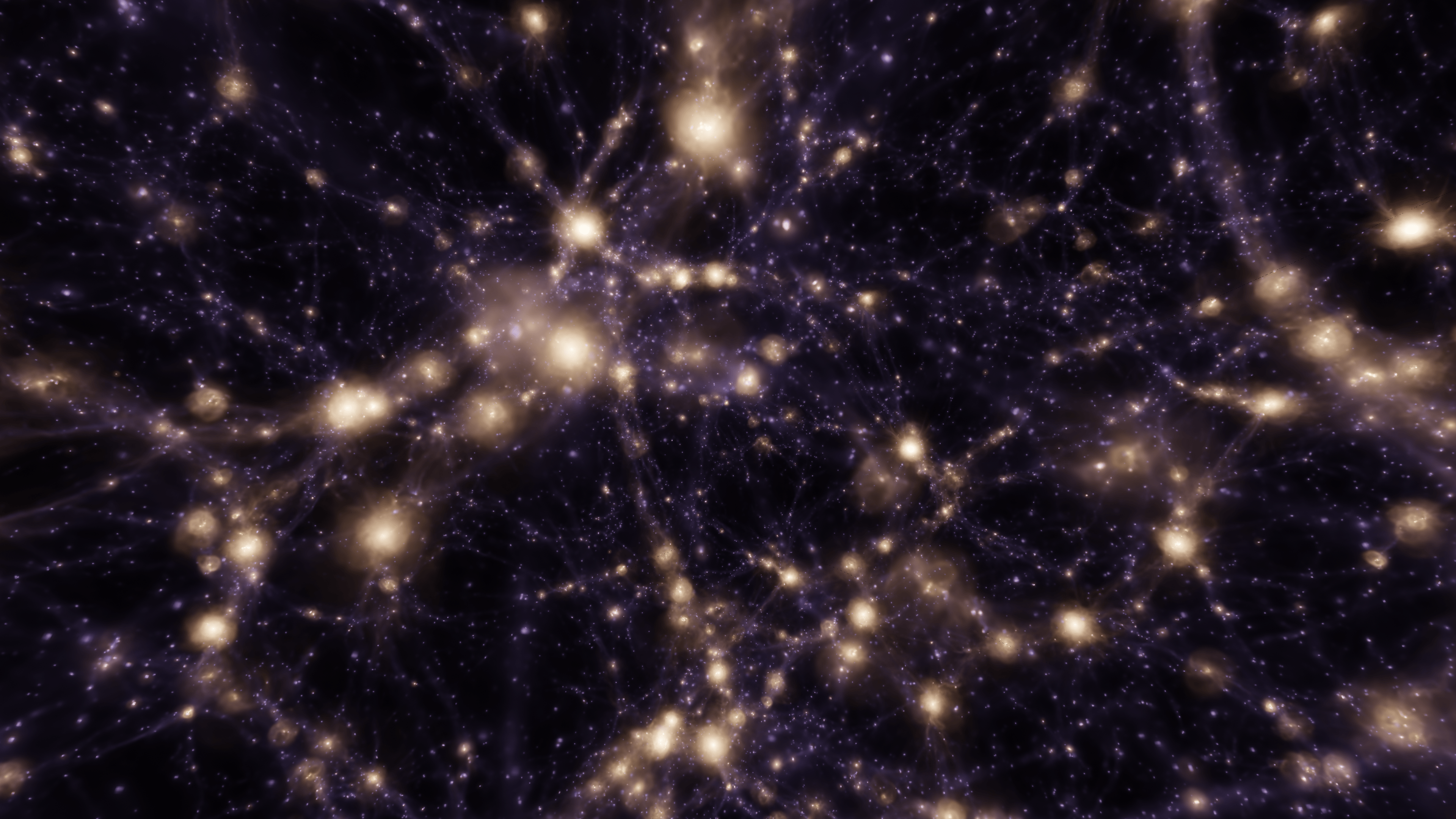}
\includegraphics[width=0.99\linewidth]{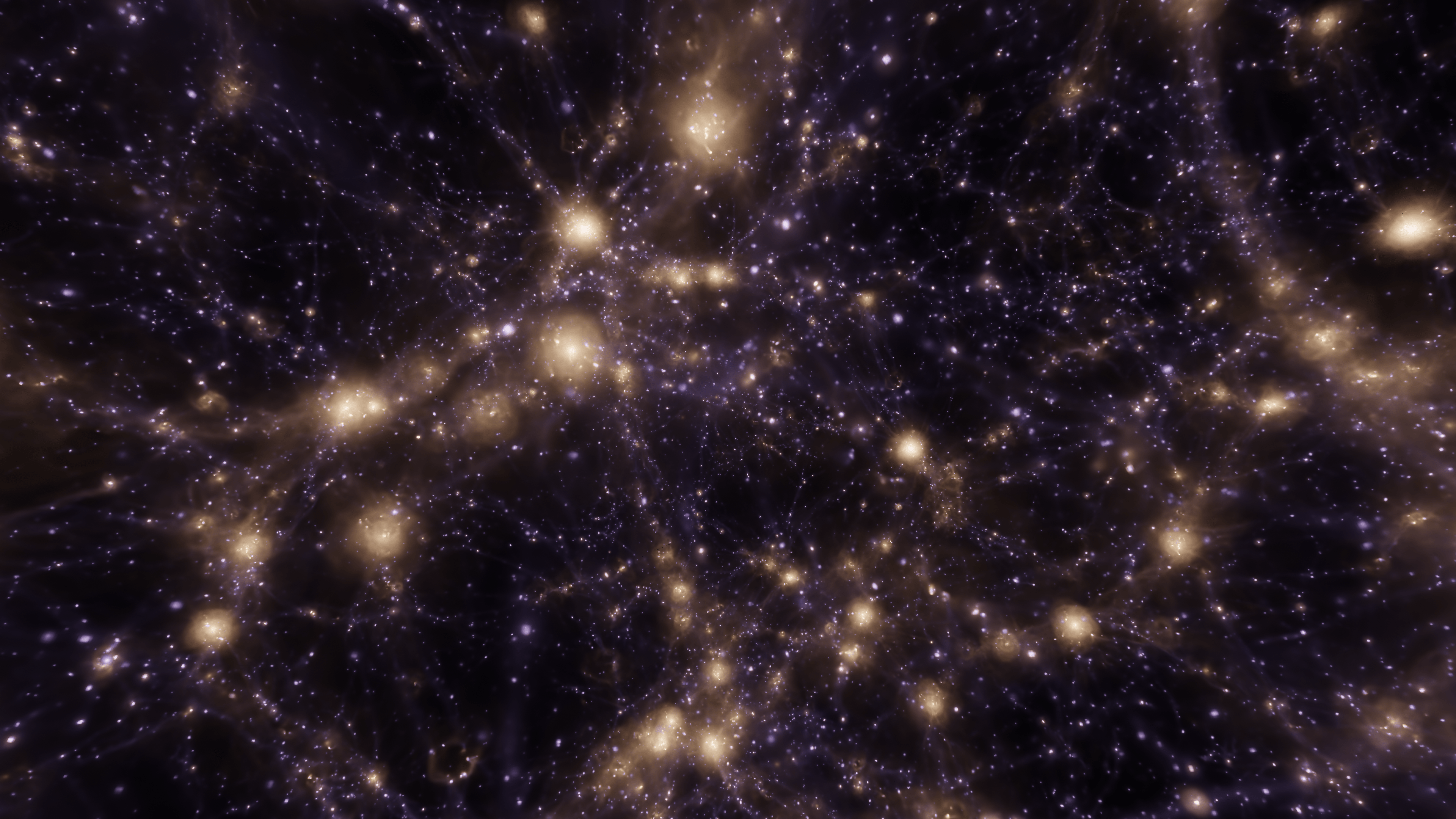}
\caption{We show two images of the gas distribution of two distinct IllustrisTNG simulations. The one on the top displays the results for a simulation with high supernova feedback strength, while the one on the bottom is from a simulation with low supernova feedback. The color represents gas temperature, while its brightness corresponds to the gas density. Finally, we apply an extinction based on gas metallicity. As can be seen, the effect of feedback is very pronounced: it not only affects the gas abundance and temperature on the smallest galaxies but it also changes the gas distribution in the most massive galaxies.}
\label{fig:renders}
\end{center}
\end{figure*}

The state-of-the-art hydrodynamic simulations have been run using two different codes, \textsc{AREPO} \citep{Arepo,Arepo_public} and \textsc{GIZMO} \citep{Hopkins2015_Gizmo}, and they made use of the IllustrisTNG \citep{WeinbergerR_16a,PillepichA_16a} and SIMBA \citep{SIMBA} galaxy formation models, respectively. However, the values of four astrophysical parameters vary from simulation to simulation. Two parameters, $A_{\rm SN1}$ and $A_{\rm SN2}$, control the efficiency of supernova feedback, while the other two parameters, $A_{\rm AGN1}$ and $A_{\rm AGN2}$, parametrize the efficiency of feedback from supermassive black holes, as described in more detail below. In Fig.~\ref{fig:renders} we illustrate visually the effect of changing one of the astrophysical parameters in one simulation. As can be seen, while the large-scale structure remains unchanged, changing the efficiency of supernova feedback has a large effect on both small and large galaxies.

For each hydrodynamic simulation, CAMELS includes its N-body counterpart. The N-body simulations have been run with \textsc{GADGET-III} \citep{gadget-3_2005}. With the simulation snapshots and initial conditions we also release the \textsc{Gadget} parameter files, \textsc{CAMB} parameters files, and linear power spectra used to run the simulations.

\subsection{Organization}

The CAMELS simulations are divided into three different suites:
\begin{itemize}
\item \textbf{IllustrisTNG}. All simulations run with the \textsc{AREPO} code and employing the IllustrisTNG model belong to this suite. There are 1,092 IllustrisTNG simulations in CAMELS.
\item \textbf{SIMBA}. All simulations run with the GIZMO code and employing the SIMBA subgrid model belong to this suite. There are 1,092 SIMBA simulations in CAMELS.
\item \textbf{N-body}. All N-body simulations belong to this suite. There are 2,049 N-body simulations in CAMELS.
\end{itemize}
We provide further details on each suite below. Each simulation suite contains four different sets, depending on the way the values of the cosmological parameters, astrophysical parameters, and initial conditions random phases are organized:
\begin{itemize}
\item \textbf{LH} stands for \textit{latin-hypercube}. This set contains 1,000 simulations, each with different values of $\Omega_{\rm m}$, $\sigma_8$, $A_{\rm SN1}$, $A_{\rm SN2}$, $A_{\rm AGN1}$, $A_{\rm AGN2}$, and the initial conditions random phases. In the case of the N-body suite, this set contains 2,000 simulations varying $\Omega_{\rm m}$, $\sigma_8$ and the initial conditions random phases, such that they match those from the IllustrisTNG and SIMBA LH sets.
\item \textbf{1P} stands for \textit{1 parameter at a time}. This set contains 61 simulations with the same values of the initial conditions random seed. The simulations only differ in the value of a single cosmological or astrophysical parameter at a time, with 11 variations for each, including the set of fiducial values. In the case of the N-body suite, this set contains 21 simulations varying $\Omega_{\rm m}$ and $\sigma_8$.
\item \textbf{CV} stands for \textit{cosmic variance}. This set contains 27 simulations that share the values of the cosmological and astrophysical parameters. The simulations only differ in the value of the initial conditions random seed. There are 27 N-body counterpart simulations for this set.
\item \textbf{EX} stands for \textit{extreme}. This set contains 4 simulations that have the same value of the initial conditions random seed and the same value of the cosmological parameters. One of them represents a model with no feedback, while the other two have either extremely large supernova or AGN feedback. The N-body suite only contains 1 simulation.
\end{itemize}

For further details on the CAMELS simulations we refer the reader to \citet{CAMELS} and references therein.

\subsection{Parameters}

Both the IllustrisTNG and SIMBA simulation suites model galaxy formation by following Newtonian gravity in an expanding background, hydrodynamics, radiative cooling, star-formation, stellar evolution and feedback, SMBH growth and AGN feedback. IllustrisTNG also follows magnetic fields in the MHD limit and SIMBA follows dust grains. The implementations of gravity and hydrodynamics solvers differ between the codes, as well as the parameterizations of radiative cooling, star-formation and stellar evolution. However, the most consequential differences between the suites are in the implementations of feedback in the form of galactic winds and from AGN, since the physics of these processes are the least theoretically understood as well as least observationally constrained. Therefore, these are also the parts of the physical modeling which we have chosen to apply variations to, through the parameters mentioned above, $A_{\rm SN1}$, $A_{\rm SN2}$, $A_{\rm AGN1}$, $A_{\rm AGN2}$, as described next.

CAMELS was designed to sample a large volume in parameter space. Thus, the value of both the cosmological and astrophysical parameters is varied within a very broad range:
\begin{eqnarray}
\Omega_{\rm m}&\in&[0.1, 0.5],\\
\sigma_8&\in&[0.6, 1.0],\\
A_{\rm SN1}&\in&[0.25, 4.0],\\
A_{\rm SN2}&\in&[0.5, 2.0],\\
A_{\rm AGN1}&\in&[0.25, 4.0],\\
A_{\rm AGN2}&\in&[0.5, 2.0].
\end{eqnarray}
In both the LH and 1P sets, the value of $\Omega_{\rm m}$ and $\sigma_8$ is sampled linearly, while the value of the astrophysical parameters is varied in logarithmic scale.

In both models, $A_{\rm SN1}$ represents a normalization factor for flux of the galactic wind feedback. In IllustrisTNG it is implemented as a pre-factor for the overall energy output per unit star-formation \citep{PillepichA_16a}, while in SIMBA it is implemented as a pre-factor for the mass-loading factor (wind mass outflow rate per unit star-formation rate) relative to that predicted by higher-resolution simulations \citep{Angles-Alcazar2017_BaryonCycle}. In both models, $A_{\rm SN2}$ represents a normalization factor for the speed of the galactic winds. This implies that for a fixed $A_{\rm SN1}$, changes in $A_{\rm SN2}$ in IllustrisTNG affect the wind speed in concert with the mass-loading factor (to keep a fixed energy output), while in SIMBA changes in $A_{\rm SN2}$ affect the wind speed in concert with the wind energy flux (with a fixed mass-loading factor).

In both models, $A_{\rm AGN1}$ represents a normalization factor for the energy output of AGN feedback while $A_{\rm AGN2}$ affects the specific energy of AGN feedback. However, the implementations of AGN feedback are quite significantly different between the suites and so is the effect of those parameters. 
In IllustrisTNG, $A_{\rm AGN1}$ is implemented as a pre-factor for the overall power injected in the `kinetic' feedback mode \citep{WeinbergerR_16a}, while in SIMBA it is implemented as a pre-factor for the momentum flux of mechanical outflows \citep{Angles-Alcazar2017_BHfeedback} in the `quasar' and `jet' feedback modes. 
In IllustrisTNG, $A_{\rm AGN2}$ directly parameterizes the burstiness and the temperature of the heated gas during AGN feedback `bursts', while in SIMBA it controls the speed of continuously-driven AGN jets. We refer the reader to \citet{CAMELS} for a detailed description of the feedback parameter variations in CAMELS.  

It is very important to remark that, in light of the discussion above, while the cosmological parameters in the N-body, IllustrisTNG, and SIMBA suites represent the very same physical effect, the astrophysical parameters in the SIMBA and IllustrisTNG suites do not. The reason is that these parameters characterize similar physical processes but in different subgrid models. Thus, one should not attempt to match these parameters across suites. In other words, when doing, e.g., parameter inference from some observable to the value of the cosmological and astrophysical parameters, and the model is trained on IllustrisTNG simulations, one can attempt to test the model to see if it is able to recover the correct cosmology from SIMBA simulations. On the other hand, one should not try to infer the value of the astrophysical parameters of IllustrisTNG simulations from a model trained on SIMBA simulations.

\begin{figure*}
\begin{center}
\includegraphics[width=0.99\linewidth]{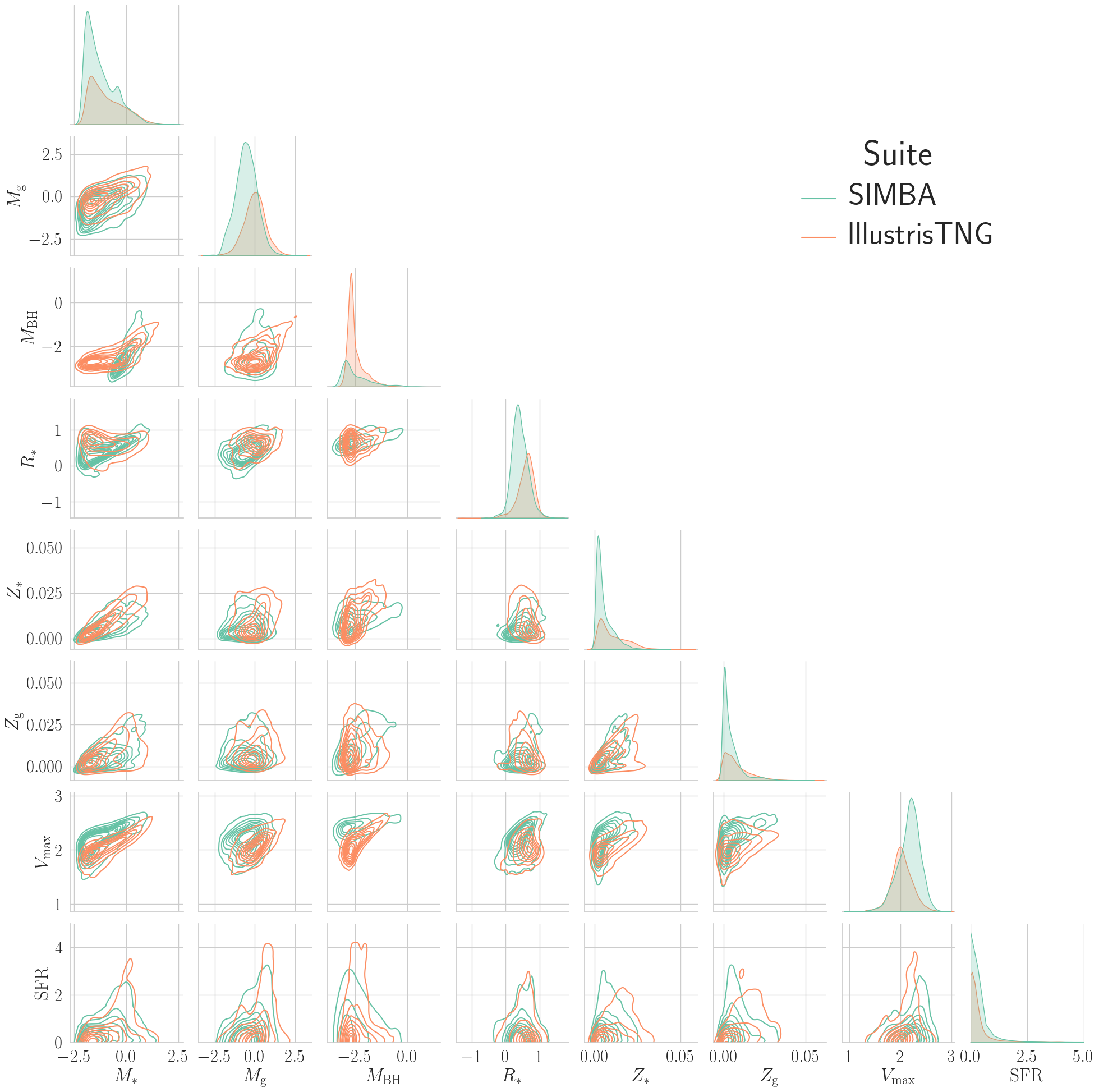}
\caption{In this plot we illustrate the similarities and differences between the IllustrisTNG and SIMBA suites considering eight different properties of the subhalos: 1) stellar mass, $M_*$, 2) gas mass, $M_{\rm g}$, 3) black-hole mass, $M_{\rm BH}$, 4) stellar half-mass radius, $R_{*}$, 5) stellar metallicity, $Z_*$, 6) gas metallicity, $Z_{\rm g}$, 7) maximum circular velocity, $V_{\rm max}$, and 8) star-formation rate, SFR. We show the 1-dimensional and 2-dimensional distribution of these properties for all galaxies in the LH sets of the IllustrisTNG (orange) and SIMBA (green) suites. Masses are in units of $10^{10}/(M_\odot/ h)$, $R_{*}$ in kpc/$h$, $V_{\rm max}$ in km/s and SFR in $M_\odot$/yr; the logarithm of each variable is shown except for the metallicity and the SFR. }
\label{fig:distributions}
\end{center}
\end{figure*}

To illustrate the differences between the IllustrisTNG and SIMBA simulations we have taken all galaxies of all simulations belonging to the LH sets of both suites. For each galaxy we consider 8 different properties and in Fig. ~\ref{fig:distributions} we show 1D and 2D distributions of them. As can be seen, while the distributions overlap in all cases, there are noticeable differences in all cases.

\section{Data Description}
\label{sec:data}

In this section we describe the different data products we release.

\subsection{Snapshots}
\label{subsec:snapshots}

We release the full snapshots generated by the \textsc{Gadget-III}, \textsc{AREPO}, and \textsc{GIZMO} codes. For each simulation, we have 34 snapshots from $z=6$ down to $z=0$ (we provide further details in the online documentation about the redshifts of the snapshots). We also release the initial conditions of each simulation.

All initial condition files and the snapshots of all simulations contain the positions, velocities, and IDs of the particles.  The snapshots of the hydrodynamic simulations contain additional fields that store properties of the gas, stars, and black-hole particles. Examples are the masses of the particles, the electron fraction from gas or the age of the star particles. We note that the simulations from the IllustrisTNG and SIMBA suites are not identical in terms of the fields they store. The differences reflect the different subgrid models employed in these two simulations. The structure and contents of the IllustrisTNG snapshots are the same as in the publicly released IllustrisTNG simulation data \citep{IllustrisTNG_public}.
Likewise, the SIMBA snapshots are identical in format to that available on the publicly released SIMBA database\footnote{\tt http://simba.roe.ac.uk}.

The snapshots are stored as \texttt{hdf5} files, and we provide details in the online documentation on how to read and manipulate the data from them.

\subsection{Halo and subhalo catalogues}

We release the halo and subhalo catalogues generated from the CAMELS simulations. The halos and subhalos have been identified using \textsc{SUBFIND} \citep{Subfind, SubfindII}, \textsc{Rockstar} \citep{Behroozi_2013}, and the Amiga Halo Finder \citep[AHF;][]{Knollmann2009_AHF}. The codes have been run on top of all snapshots of all simulations. In total, we release 431,766 catalogues that contain millions of halos, subhalos, and galaxies. We now describe the catalogues in more detail.

\subsubsection{Subfind}
\label{subsec:subfind}
\textsc{Subfind} \citep{Subfind, SubfindII} was run on-the-fly for the IllustrisTNG simulations while for the SIMBA and N-body simulations it was run in post-processing. \textsc{Subfind} identifies both halos and subhalos, and computes several physical quantities for them in both the N-body and hydrodynamic simulations. We release all \textsc{Subfind} catalogues (one per simulation and redshift) for all simulations and all redshifts. The data is stored as \texttt{hdf5} files, and we provide details on how to read the data and the information stored in them in the online documentation.

\subsubsection{Amiga Halo Finder}
\label{subsec:ahf}
AHF \citep{Knollmann2009_AHF} was run in post-processing for the IllustrisTNG and SIMBA simulations. AHF utilizes isodensity contours to locate halo centers. Halo virial radii are defined to represent spherical overdensity regions with 200 times the critical density. We release the AHF catalogues for all simulations and redshift snapshots, including (1) global halo properties, (2) radial profiles, and (3) particle ID lists to identify the host halo of each particle.  We provide further details on the format and how to read these catalogues in the online documentation.

\subsubsection{Rockstar}
\label{subsec:rockstar}
In addition to the {\sc Subfind} and AHF halo catalogs, we also release halo
catalogs constructed using the {\sc Rockstar} halo finder~\citep{Behroozi_2013}. 
{\sc Rockstar} identifies dark matter halos based on an adaptive hierarchical
refinement of friends-of-friends in six-dimensional phase space plus time. 
Substructures are identified using successively smaller linking length and
particles are assigned to the inner-most substructures, which are defined 
as halo seeds, based on their phase-space proximity. 
We release {\sc Rockstar} halo catalogs of all 34 snapshots from $z=6$ to 0 for all simulations.

Furthermore, we use {\sc Consistent-Trees}~\citep{behroozi_2013b} to
generate merger trees from the {\sc Rockstar} halo catalogs.  
We note that {\sc Consistent-Trees} ensures consistency of halo mass, 
position, and velocity across time steps.
Since all CAMELS simulations have only 34 snapshots, we perform the following exercise to quantify its validity. We have compared the {\sc Rockstar} + {\sc Consistent-Trees} outputs at $z=0$ 
from two CAMELS simulations that have the same initial conditions
but different time resolution (34 versus 200 snapshots).
We find good agreement between the outputs for certain proxies of
merger history such as peak mass and half-mass assembly time.
However, we caution readers when using more detailed properties of the 
halo merger histories, such as accretion history, which are affected
by the lower time sampling. 
All {\sc Rockstar} catalogues and {\sc Consistent-Trees} merger trees 
occupy 1.2 Terabytes of data.

\subsection{Void catalogues}
\label{subsec:VIDE}

We release void catalogs built from the CAMELS simulations with the Void IDentification and Examination toolkit (\texttt{VIDE}) \cite{Sutter_2014VIDE}. \texttt{VIDE}, based on \texttt{ZOBOV} \citep{vide:Neyrinck-2008}, has been widely used to find voids both in data---e.g. voids from the SDSS BOSS \citep{Hamaus_2016, Hamaus_2020} and eBOSS \citep{Aubert_2020} datasets, or data from DES \citep{Pollina_2018}---and simulations \citep[e.g.][]{Kreisch_2019, verza_2019,Contarini_2020, Kreisch_2021}. Furthermore, \texttt{VIDE} has also been applied to hydrodynamic simulations \citep{Habouzit_2020,Panchal_2020}, showing its suitability for the CAMELS dataset. 

\texttt{VIDE} was run on top of CAMELS galaxies, that were defined as subhalos containing more than 20 star particles. Given the size of the CAMELS simulations, and the extended size of cosmic voids (that usually span sizes from $5-100~ h^{-1}\mathrm{Mpc}$), the number of voids for each CAMELS simulation is relatively small. The \texttt{VIDE} catalogues store information about the positions, sizes, ellipticities, and member galaxies of each void. In the online documentation we provide further details on how to read and manipulate the \texttt{VIDE} catalogues.

\subsection{Lyman-alpha spectra}
\label{subsec:Lya}

We release mock Lyman-$\alpha$ spectra generated using a public, well-tested code exhibited in \citet{Bird+2015} and used previously for studies of the Lyman-$\alpha$ Forest in \citet{Gurvich_2017}. The spectra is generated for 5,000 sightlines randomly placed through the simulation box. This spectral data was generated for the IllustrisTNG and SIMBA suites for all simulation sets at all redshifts. The locations of the random sightlines vary across snapshots.

The total absorption along a sightline is the sum of the absorption from all the nearby gas cells. The simulated spectra has a spectral resolution of 1 km/s and any lines with an optical depth of $\tau < 10^{-5}$ are neglected. For further details on the artificial spectra calculation, we direct the reader to \citet{Bird+2015}. In the online documentation we provide details on how to read and manipulate the Lyman-$\alpha$ spectra.

\vspace{1cm}
\subsection{Summary statistics}

We release a large set of summary statistics, containing power spectra, bispectra, and probability distribution functions. This data can be used for a large variety of tasks such as carrying out parameter inference and building emulators.

\subsubsection{Power spectrum}
\label{subsubsec:pk}

The power spectrum is the most prominent summary statistic of cosmology. The procedure used to carry out this task is the following. First, the positions and masses of the considered particles are read from the snapshots. Next, the masses of the particles are deposited into a regular grid with $512^3$ voxels using the Cloud-in-Cell mass-assignment scheme (MAS). We then Fourier transform that field and correct modes amplitudes to account for the MAS. Finally, the power spectrum is computed by averaging the square of the modes amplitudes 
\begin{equation}
    P(k_i)=\frac{1}{N_i}\sum_{k\in k_{\rm bin}}\left|\delta(\mathbf{k})\right|^2,
\end{equation}
where the $k$-bins have a width equal to the fundamental frequency, $k_F=2\pi/L$ ($L$ is the box size), and $N_i$ is the number of modes in the $k$-bin. The wavenumber associated with each bin is
\begin{equation}
    k_i = \frac{1}{N_i}\sum_{k\in k_{\rm bin}}k~.
\end{equation}
We have computed the power spectra of the total matter for both the N-body and the hydrodynamic simulations. Besides, for the hydrodynamic simulations we have also computed the power spectra of the gas, dark matter, stars, and black hole components. We have done this for all snapshots of each simulation. We have made use of \textsc{Pylians}\footnote{\url{https://pylians3.readthedocs.io}} to carry out the calculation.  In total, we release 440,946 power spectra. All power spectra occupy $\simeq 10$ Gigabytes of data.

The above methods are inefficient if we wish to compute the power spectrum at large $k$, since they require a unwieldy FFT grid. In this regime, alternative methods such as configuration-space power spectrum estimators \citep{philcox_eisenstein_bk20} can be of use, since their computational cost decreases as the minimum scale increases. We provide power spectrum multipoles computed up to $k = 1,000~h\,\mathrm{Mpc}^{-1}$ and $\ell_{\rm max} = 4$, using a combination of the above \textsc{Pylians} code and the \textsc{hipster} pair-counting approach package \citep{philcox_bk20}, switching between the two at $k  = 25~h\,\mathrm{Mpc}^{-1}$ and convolving the small-scale spectra with a window of size $R_0 = 1~h^{-1}\mathrm{Mpc}$ for efficiency. Spectra are computed at $z=0$ for all matter species listed above, and we include results from each simulation of the LH set from the IllustrisTNG, SIMBA, and N-body suites in both real- and redshift-space, with the latter using three choices of redshift-space axis. In total we compute 44,000 power spectra up to $k = 1,000~h\,\mathrm{Mpc}^{-1}$, requiring $\simeq 14,000$ CPU-hours and occupying $\simeq 0.6$ Gigabytes of storage.

For all spectra, we store the value of $k$ in each k-bin, the value of $P(k_i)$, and (for the large-scale spectra) the number of modes in each bin. We provide further details on how to read and manipulate these files in the online documentation.

\subsubsection{Bispectrum}
\label{subsubsec:bk}

On large scales, the first non-Gaussian statistic of interest is the bispectrum, encoding the three-point average of the density field. In this release, we provide bispectrum measurements from gas, dark matter and total matter for the 1,000 simulations of the LH set of the IllustrisTNG and SIMBA suites, as well as 1,000 N-body simulations. These are performed at redshift zero, both in real-space and redshift-space (for three choices of line-of-sight). Additional data can be computed upon request.

On large scales, bispectra are computed analogously to \S\ref{subsubsec:pk}, first gridding the data with $128^3$ voxels using a Triangular-Shaped-Cloud MAS scheme. We then use the \textsc{Pylians} estimator \citep{pylians}, implementing the approach of \citet{watkinson17}, which practically computes the following sum via a series of FFTs:
\begin{equation}\label{eq: bk-computation}
    B(k_1,k_2,\mu) = \frac{\sum_{\mathbf{k}_1}\sum_{\mathbf{k}_2}\delta(\mathbf{k}_1)\delta(\mathbf{k}_2)\delta(-\mathbf{k}_1-\mathbf{k}_2)}{N_T(k_1,k_2,\mu)}.
\end{equation}
The bispectrum is parametrized by two lengths, $k_1$ and $k_2$, and an internal angle $\mu\equiv \hat{\mathbf{k}}_1\cdot\hat{\mathbf{k}}_2$, with $N_T(k_1,k_2,\mu)$ giving the number of triangles per bin. We use $20$ $k$-bins with $\Delta k = 0.25~h\,\mathrm{Mpc}^{-1} \approx k_F$, and ten linearly spaced $\mu$ bins.

The above method becomes prohibitively expensive as $k_{\rm max}$ (and thus the FFT grid) increases. To ameliorate this, we compute the bispectra at large $k$ using the \textsc{hipster} code, as for the small-scale power spectra, here convolving the spectra with a smooth window of scale $R_0=2~h^{-1}\mathrm{Mpc}$. This computes the Legendre multipoles of the bispectrum, related to Eq.\,\ref{eq: bk-computation} by
\begin{equation}
    B(k_1,k_2,\mu) = \sum_{\ell=0}^\infty B_\ell(k_1,k_2)L_\ell(\mu),
\end{equation}
for Legendre polynomial $L_\ell(\mu)$, and uses 25 linearly spaced $k$-bins in the range $[0,50]h\,\mathrm{Mpc}^{-1}$ for $\ell\leq 5$, subsampling to $10^5$ particles for efficiency. These bispectra are computed for the same simulations as before, and will allow information to be extracted from very small scales. In total, 28,000 bispectra are estimated using each method, requiring $\simeq$ 70,000 CPU-hours and $\simeq 2.1$ Gigabytes of storage.

\subsubsection{Probability distribution function}
\label{subsec:PDF}

We estimate probability distribution functions (PDF) for 13 different physical fields using the 3D grids of the CAMELS Multifield Dataset (CMD) (see Sec. \ref{sec:CMD}). The PDFs are calculated for all the fields: 1) gas temperature, 2) gas pressure, 3) neutral hydrogen density, 4) electron number density, 5) gas metallicity, 6) gas density, 7) dark matter density,  8) total mass density, 9) stellar mass density, 10) magnetic fields, 11)  ratio between magnesium over iron, 12) gas velocity, and 13) dark matter velocity, for all the grid sizes, i.e., $128$, $256$ and $512$ at redshifts $0.0$, $0.5$, $1.0$, $1.5$, and $2.0$. The PDFs are calculated as follows. First, the 1,000 3D grids from all simulations in the LH set are read into memory. We then calculate the minimum value across grids and if it equals $0$, a small offset is added to all voxels of all grids. The offset, $\varepsilon$, is given by

\begin{equation}
    \varepsilon = \frac{\rm{min_{non-zero}}}{10^{20}},
    \label{eqn_pdf}
\end{equation}
where $\rm{min_{non-zero}}$ denotes the non-zero minimum of all the 1,000 grids. Then we log-transform the entire field (to the base 10) and construct a histogram of $500$ bins between the minimum and maximum values of the entire field. Finally, we save to disk the number of counts in each bin for each grid in the considered field.

\subsection{Profiles}
\label{subsec:profiles}

We provide three-dimensional spherically-averaged profiles of gas density, thermal pressure, gas mass-weighted temperature, and gas mass-weighted metallicity for the 1P, LH, and CV sets of both the IllustrisTNG and SIMBA suites.  We follow \citet{Moser2021} in extracting halo information and construction of the profiles. Specifically, we use  \texttt{illstack\char`_CAMELS}\footnote{\url{https://github.com/emilymmoser/illstack_CAMELS}} (a CAMELS-specific version of the original, more general code \texttt{illstack} used in \citealt{Moser2021}), to generate the three-dimensional profiles, extending radially from $0.01-10$ Mpc in 25 $\log_{10}$ bins. The profiles are stored in \texttt{hdf5} format which can be read with the python script provided in the \texttt{illstack\char`_CAMELS} repository.

\vspace{0.5cm}
\subsection{X-rays}
\label{subsec:Xrays}
   
We provide mock X-ray photon lists in the form of SIMPUT fits files for all halos above $10^{12}\;M_{\odot}$ across all hydrodynamic CAMELS runs at redshift $z=0.05$ obtained from the snapshot 032.  The SIMPUT files are generated using the pyXSIM package\footnote{\url{http://hea-www.cfa.harvard.edu/~jzuhone/pyxsim/}} and contain positional coordinates in RA and DEC coordinates and energy in units of keV.  These files serve as inputs into other software packages, including SOXS\footnote{\url{http://hea-www.cfa.harvard.edu/~jzuhone/soxs/}} and SIXTE \citep{sixte} that generate mock observations for specific telescopes using custom instrument profiles. These SIMPUT files can also represent idealized observations by an X-ray telescope, and we also provide a single collated file with 1-dimensional projected surface brightness (SB) profiles for all halos for the soft X-ray band (0.5-2.0 keV) in units of erg s$^{-1}$ kpc$^{-2}$.  This file holds 160,693  SB profiles across the 2,190 1P, CV, LH, and EX simulations. 

\subsection{CAMELS Multifield Dataset}
\label{sec:CMD}

The CAMELS Multifield Dataset, CMD, is a collection of hundreds of thousands of 2D maps and 3D grids generated from CAMELS data. CMD contains 15,000 2D maps for 13 different fields at $z=0$, and 15,000 3D grids, at three different spatial resolutions and at five different redshifts. The data was generated by assigning particles positions and properties (e.g. mass and temperature for the temperature field) to either 2D maps or 3D grids. There are many possible machine learning applications of this dataset, e.g.: 1) parameter inference \citep{Paco_2021a,Paco_2021b}, 2) summary or field level emulation, 3) mapping N-body to hydrodynamic simulations, 4) superresolution, and 5) time evolution. In total, CDM represents over 70 Terabytes of data. We refer the reader to \cite{Paco_2021c} and the CMD online documentation\footnote{\url{https://camels-multifield-dataset.readthedocs.io}} for further details on this dataset.

\subsection{CAMELS-SAM}
\label{subsec:SAM}

CAMELS-SAM represents a newer third `hump' of CAMELS, mimicking its construction and purpose but using larger N-body volumes that are populated with galaxies using the Santa Cruz semi-analytic model (SAM, \citealt{Somerville2008, Somerville2015}) of galaxy formation. The N-body simulations are run with \textsc{AREPO} \citep{Arepo_public}, and follow the evolution of 640$^3$ dark matter particles over a periodic box of (100 h$^{-1}$ cMpc)$^3$ volume from $z=127$ to $z=0$. For each simulation we save 100 snapshots. The initial conditions were otherwise generated as described in \textsection \ref{sec:simulations}, with the same underlying cosmology, and a newly generated latin hypercube varying $\Omega_m$, $\sigma_8$, and three SAM parameters. Those parameters were chosen as the ones closest to the astrophysical parameters varied in CAMELS. Two parameters control the amplitude and rate of mass outflow from massive stars out of a galaxy, and the third parameter broadly controlling the strength of the radio jet mode of AGN. 

Like CAMELS, CAMELS-SAM has an LH set containing 1,000 simulations. The values of the cosmological and astrophysical parameters in the set are organized in a latin-hypercube. We additionally have 5 simulations in the CV set where the value of the initial random seed varies and the 5 parameters are held fixed to their fiducial values. Finally, a 1P set with 12 simulations exists, where the SC-SAM was run at the smallest and largest value of each SAM parameter for two of the CV simulations. 

It is important to emphasize the differences between the original CAMELS and the CAMELS-SAM simulations. First, CAMELS-SAM consists of N-body simulations with a volume $64\times$ larger than the former, while CAMELS contains both N-body and hydrodynamic simulations. Second, CAMELS-SAM stored 100 snapshots while CAMELS only kept 34. Third, galaxies are modelled in very different ways: in CAMELS they arise from the hydrodynamic simulations while in CAMELS-SAM they are modelled through the Santa Cruz semi-analytic model.

For all CAMELS-SAM simulations, we release:
\begin{itemize}
\item The halo and subhalo catalogues from both \textsc{Subfind} and \textsc{Rockstar}.
\item The merger trees generated from \textsc{Consistent trees}.
\item The galaxy catalogues from the Santa Cruz SAM.
\end{itemize}

The galaxy catalogues are stored as \textit{.dat} text files with comma-separated values. These files contain information about the halo and galaxies from all snapshots of a given simulation. The exact available properties, their organization and units, and example code to open these files can be found on the CAMELS-SAM online documentation\footnote{\url{https://camels-sam.readthedocs.io}}. The total size of these data products is around 50 Terabytes. 

The raw data (compressing full N-body snapshots across redshifts) has been stored on tape and its content can be retrieved upon request. We refer the reader to \textcolor{blue}{Perez, Genel, et al. (2022)} for further details on CAMELS-SAM, as well as a proof-of-concept of its power using clustering summary statistics to constrain cosmology and astrophysics with neural networks.

\section{Data Access and structure}
\label{sec:access}

In this section we describe the different methods to access the data and its structure.

\subsection{Data Access}

We provide access to CAMELS data through four different platforms:

\begin{figure*}
\begin{center}
\includegraphics[width=0.99\linewidth]{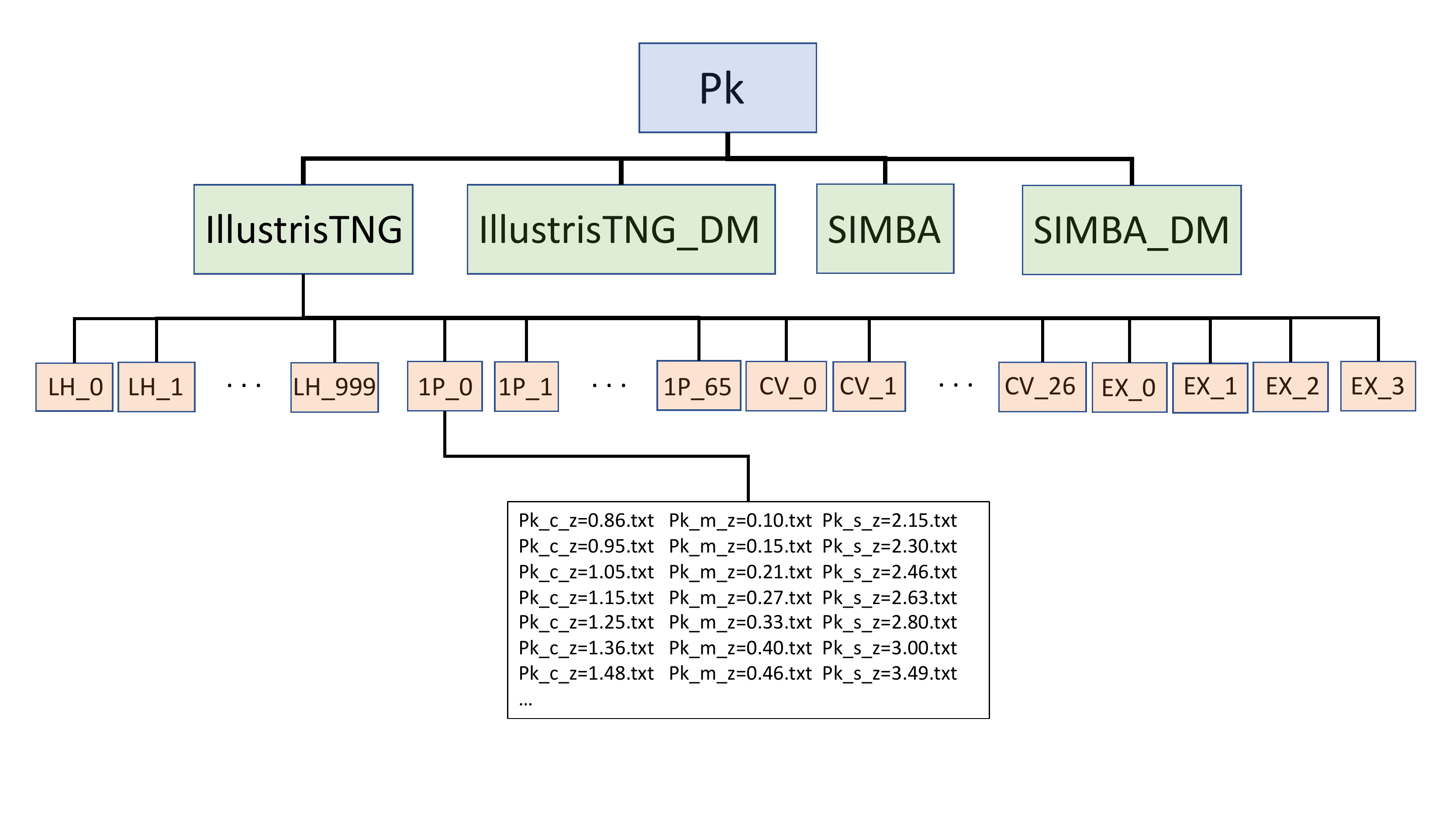}
\caption{This scheme shows the generic structure of CAMELS data. The top level represents the type of data it contains (power spectra in this case). Inside that folder there are typically four folders containing the data for the three different simulation suites: IllustrisTNG, SIMBA, and their N-body counterparts (\texttt{IllustrisTNG\_DM} and \texttt{SIMBA\_DM}). Within each of those folders there are numerous folders, containing the data from the different simulations belonging to each suite; i.e. the simulations from the four sets: LH, 1P, CV, and EX. Finally, inside each of those folders the user can find the data products themselves. In this particular case, the power spectra for the different component.}
\label{fig:scheme}
\end{center}
\end{figure*}

\begin{itemize}
\item \textbf{Binder}. Binder is a system that allows users to read and manipulate data that is hosted at the Flatiron Institute through either a Jupyter notebook or a unix shell. The system provides access to the entire CAMELS data and allows users to perform calculations that do not require large amounts of CPU power. We note that heavy calculations are not supported by this system, and we recommend the user to download the data locally and work with it accordingly. We provide the link to the Binder environment in the online documentation. All CAMELS data can be accessed, read, and manipulated through Binder. 
We provide further technical details on Binder usage in the online documentation.
\item \textbf{Globus}. Globus\footnote{\url{https://www.globus.org}} is a system designed to transfer large amounts of data in an efficient way. All CAMELS data can be transferred through globus. We provide the globus link in the online documentation\footnote{Since this link may change with time, we make it available in the online documentation, where it can be updated if needed.}. Users can transfer the data to either another cluster or directly to their personal computer.
\item \textbf{Url}. We also provide a uniform resource locator (url) to access the data through a browser. We do not recommend transferring large quantities of data using this procedure, as both the speed and its reliability is much worse than globus. On the other hand, to download small amounts of data, such as a particular power spectrum or a halo catalogue, it may be useful. All CAMELS data can be accessed and downloaded through the url. We provide the url link in the online documentation where it will be always updated.
\item \textbf{FlatHUB}. FlatHUB is a platform that allows users to explore and compare data from different simulations by browsing and filtering the data, making simple preview plots, and downloading sub-samples of the data. We provide access to the \textsc{Subfind} halo and subhalo catalogues of the IllustrisTNG and SIMBA suites through this platform. We provide a link to FlatHUB in the online documentation.
\end{itemize}

\subsection{Data Structure}

The data is organized in different folders that contain similar type of data:
\begin{itemize}
\item \textbf{Sims}. This folder contains the raw data from the simulations, such as initial conditions, snapshots, and parameter files. This folder contains 205 terabytes of data.
\item \textbf{FOF\_Subfind}. This folder contains the \textsc{SUBFIND} halo and subhalo catalogues described in Sec. \ref{subsec:subfind}. This folder contains 4 terabytes of data.
\item \textbf{AHF}. This folder contains the AHF halo catalogues described in Sec.~\ref{subsec:ahf}. This folder contains 6 terabytes of data.
\item \textbf{Rockstar}. This folder contains the \textsc{Rockstar} halo and subhalo catalogues together with the merger trees from \textsc{Consistent-trees} as described in Sec. \ref{subsec:rockstar}. This folder contains 1 terabyte of data.
\item \textbf{Pk}. This folder contains the power spectra described in Sec. \ref{subsubsec:pk}. This folder contains approximately 10 gigabytes of data.
\item \textbf{Bk}. This folder contains the bispectra measurements described in Sec. \ref{subsubsec:bk}. This folder contains approximately 2.6 gigabytes of data.
\item \textbf{CMD}. This folder contains the CAMELS Multifield Dataset. This folder contains 76 terabytes of data.
\item \textbf{VIDE\_Voids}. This folder contains the void catalogues described in Sec. \ref{subsec:VIDE}. This folder contains 200 megabytes of data.
\item \textbf{Lya}. This folder contains the Lyman-$\alpha$ spectra described in Sec. \ref{subsec:Lya}. This folder contains 14 terabytes of data.
\item \textbf{PDF}. This folder contains the probability distribution function measurements described in Sec. \ref{subsec:PDF}. This folder contains more than 1 gigabyte of data.
\item \textbf{Profiles.} This folder contains the spherically-averaged 3D profiles described in Sec. \ref{subsec:profiles}. This folder contains 48 gigabytes of data.
\item \textbf{X-rays}. This folder contains the X-rays photon lists described in Sec. \ref{subsec:Xrays}. This folder contains over 100 gigabytes of data.
\item \textbf{SCSAM}. This folder contains all CAMELS-SAM data products described in Sec. \ref{subsec:SAM}. This folder contains more than 50 terabytes.
\item \textbf{Utils}. This folder contains additional files that can be useful to the user, including a file with the value of the scale factors corresponding to simulation snapshots and files indicating the values of the cosmological and astrophysical parameters of each simulation.

\end{itemize}

When possible, we have organized the data in the different folders in a self-similar way. We show the generic data structure scheme in Fig. \ref{fig:scheme}. The data is first organized into folders that contain: 1) the IllustrisTNG hydrodynamic simulations, 2) the SIMBA hydrodynamic simulations, 3) the N-body counterparts of 1), and 4) the N-body counterparts of 2). Inside each of these folders the user can find many different sub-folders whose name refers to the specific simulation set and realization: e.g. the first simulation of the LH set is denoted as LH\_0. Finally, inside each of those folders the user can find the data with the particular characteristics of each data product. We note that these folders may contain data products for a particular CAMELS simulation at all redshifts.

For some data products, e.g. CMD and CAMELS-SAM, the data organization is slightly different to the one outlined above. In those cases, we provide further details in the online documentation.

\section{Summary}
\label{sec:summary}

The goal of the CAMELS project is to connect cosmology with astrophysics via thousands of state-of-the-art cosmological hydrodynamic simulations and extract the maximum amount of information from them via machine learning. CAMELS contains 4,233 cosmological simulations, 2,049 N-body simulations and 2,184 state-of-the-art hydrodynamic simulations sampling a vast volume in parameter space using two independent codes that solve hydrodynamic equations and implement subgrid physics in very distinct ways. CAMELS data have already been used for a large variety of tasks, from providing the first constraints on the mass of the Milky Way and Andromeda galaxies using artificial intelligence to showing that neural networks can extract information from vastly different physical fields while marginalizing over astrophysical effects at the field level.

In this paper we have described the characteristics of the CAMELS simulations and a variety of additional data generated from them, including halo, subhalo, galaxy, and void catalogues, power spectra, bispectra, Lyman-$\alpha$ spectra, probability distribution functions, radial profiles, and X-rays photon lists. We have also described CAMELS-SAM, a collection of more than 1,000 galaxy catalogues created by applying the Santa Cruz Semi-Analytic Model to a set of hundreds of N-body simulations. We have made all this data publicly available, comprising hundreds of terabytes. We provide access to the data through different platforms, including a Binder environment for interactive data manipulation with Jupyter notebooks, a Globus link for efficient transfer of large amounts of data, and the FlatHUB platform for quick exploration of  \textsc{Subfind} (sub)halo catalogues. We emphasize that the information outlined in this paper may become outdated as additional data products become available over time. However, the online documentation located at \url{https://camels.readthedocs.io} will always be updated accordingly. 

It is also important to be aware of the limitations associated to the CAMELS simulations. First, the volume sampled by each individual simulation is relatively small, $(25~h^{-1}{\rm Mpc})^3$, inhibiting the formation of the most extreme objects in the Universe such as galaxy clusters and large voids. Second, while CAMELS covers a large volume in parameter space, it would be desirable to make it even larger by including other cosmological and astrophysical parameters. Third, CAMELS only contains two distinct suites of hydrodynamic simulations: IllustrisTNG and SIMBA. Ideally, we would like to expand CAMELS to simulations performed with additional codes employing different subgrid models. Fourth, the resolution of CAMELS may not be high enough for some astrophysical problems. Future versions of CAMELS will be designed to tackle these limitations.

We believe that CAMELS data will become a powerful tool for the community. 

\section*{ACKNOWLEDGEMENTS}
We are indebted to the high-performance computing system administrators at the Flatiron Institute and Princeton University for their invaluable help and work accommodating the CAMELS storage needs. The authors are pleased to acknowledge that the work reported in this paper was partially performed using the Research Computing resources at Princeton University which is a consortium of groups led by the Princeton Institute for Computational Science and Engineering (PICSciE) and the Office of Information Technology's Research Computing Division. Further technical details on the CAMELS simulations and instructions to download the data can be found in \url{https://camels.readthedocs.io} and \url{https://www.camel-simulations.org}. The work of FVN, SG, DAA, SH, OP, AP, KW, WC, ME, US, DS, BB, BW, RS, and GB was supported by the Simons Foundation. DAA was supported in part by NSF grants AST-2009687 and AST-2108944. BB is grateful for generous support by the David and Lucile Packard Foundation and Alfred P. Sloan Foundation. The Flatiron Institute is supported by the Simons Foundation.

\bibliography{references}{}
\bibliographystyle{aasjournal}

\end{document}